\documentclass[12pt,twoside]{article}   
\usepackage{epsfig,times}  
\usepackage{color}

\begin{document}
\thispagestyle{empty} 


 \renewcommand{\topfraction}{.99}      
 \renewcommand{\bottomfraction}{.99} 
 \renewcommand{\textfraction}{.0}


\newcommand{\nc}{\newcommand}

\nc{\qI}[1]{\section{{#1}}}
\nc{\qA}[1]{\subsection{{#1}}}
\nc{\qun}[1]{\subsubsection{{#1}}}
\nc{\qa}[1]{\paragraph{{#1}}}

\def\qbu{\hfill \par \hskip 6mm $ \bullet $ \hskip 2mm}
\def\qee#1{\hfill \par \hskip 6mm #1 \hskip 2 mm}

\nc{\qfoot}[1]{\footnote{{#1}}}
\def\qL{\hfill \break}
\def\qpar{\vskip 2mm plus 0.2mm minus 0.2mm}
\def\tvi{\vrule height 12pt depth 5pt width 0pt}
\def\qtvi{\vrule height 2pt depth 5pt width 0pt}
\def\qth{\vrule height 15pt depth 0pt width 0pt}
\def\qtb{\vrule height 0pt depth 5pt width 0pt}

\def\qparr{ \vskip 1.0mm plus 0.2mm minus 0.2mm \hangindent=10mm
\hangafter=1}

\def\qdec#1{\par {\leftskip=2cm {#1} \par}}
%
\def\qbfb#1{{\bf\color{\blue}{#1} }}

\def\qdpt{\partial_t}
\def\qdpx{\partial_x}
\def\qddpt{\partial^{2}_{t^2}}
\def\qddpx{\partial^{2}_{x^2}}
\def\qn#1{\eqno \hbox{(#1)}}
\def\qds{\displaystyle}
\def\qw{\widetilde}
\def\qmax{\mathop{\rm Max}}   
\def\qmin{\mathop{\rm Min}}   

\def\qs#1{{\bf \color{blue} \LARGE {#1}}\quad }

\def\qv{\vskip 0.1mm plus 0.05mm minus 0.05mm}
\def\qhu{\hskip 1mm}
\def\qhv{\hskip 3mm}
\def\qvv{\vskip 0.5mm plus 0.2mm minus 0.2mm}
\def\qhw{\hskip 1.5mm}
\def\qleg#1#2#3{\noindent {\bf \small #1\qhw}{\small #2\qhw}{\it \small #3}\qv }


\centerline{\bf \Large  Is there an infant mortality in bacteria?}
\vskip 5mm


\centerline{Eduardo M. Garcia-Roger$ ^1 $,
Peter Richmond$ ^2 $, Bertrand M. Roehner$ ^3 $}

\vskip 10mm


\centerline{Version of 13 March 2021, provisional, comments are welcome}

\vskip 10mm

{\small Key-words: Bacteria, division, death, flow cytometry, 
congenital anomalies}

\vskip 5mm

{\bf Abstract} \qL
This manuscript proposes a significant step in our
long-run investigation 
of infant mortality across species. Since 2016
(Berrut et al. 2016) 
a succession of studies ( Bois et al. 2019)
have traced infant
mortality from organisms of high complexity (e.g. mammals)
down to unicellular organisms. \qL
Infant mortality may be considered as
a filtering process through which organisms
with potentially lethal congenital defects 
are eliminated. Such defects may
have many causes but here we focus particularly
on mishaps resulting from non-optimal conditions
in the production of proteins, enzymes and other
crucial macromolecules. \qL
The 
statistical signature of infant mortality
consists in a falling
age-specific death rate. \qL
The question we
address here is whether infant 
mortality episodes take place
in bacteria in the minutes precededing or following cell division. 
It will be shown that while experiments carried out in the
20th century tried but failed to detect such an effect (mostly because
of limited sample size), more recent
observations provided consistent evidence of
a sizeable mortality, with a rate of the order of 0.7 per 1,000
per hour, in the exponential growth phase of
{\it E. coli}. A further crucial test will be to measure the
age-specific, post-division death rate. An experiment 
is outlined for that purpose. 
It is based on the selection of stained
cells through flow cytometry and the derivation of their ages
at death from their sizes.\qL
If an infant mortality effect can be identified in {\it E. coli}
it can be conjectured that a similar effect also exists
in other unicellular organisms, both prokaryote and
eukaryote.

\vskip 10mm

\count101=0  \ifnum\count101=1

1: Institut National de Recherche pour l'Agriculture, l'Alimentation
et l'Environnement (INRAE).\qL
Email: arnaud.chastanet@inrae.fr
\qpar

\fi

1: Institut Cavanilles de Biodiversitat I Biologia Evolutiva,
University of Valencia, Spain. .\qL
Email: eduardo.garcia@uv.es
.\qpar

2: School of Physics, Trinity College Dublin, Ireland. \qL
Email: peter\_richmond@ymail.com
\qpar

3: Institute for Theoretical and High Energy Physics (LPTHE),
Pierre and Marie Curie Campus, Sorbonne University, National
Center for Scientific Research (CNRS), Paris, France. \qL
Email: roehner@lpthe.jussieu.fr

\vfill \eject

 \qdec{The juxtaposition in the title of the expressions 
``infant mortality'' along with ``bacteria'' may at first appear
puzzling. However, recall that (i) ``infant mortality''
is a notion also used in reliability science and
(ii) that the question raised in the title
has already been investigated (in these very terms) 
89 years ago (Kelly et al. 1932).
Yet, it is only in the past two decades
that some preliminary answers have slowly started to emerge (see below).
The specific definition of infant mortality to which we refer
is a phase following birth 
during which the age-specific mortality is a
{\it decreasing} function of age.}

\qI{Introduction}

Within the architectural master plan provided by the genome,
the information stored on each gene is
to serve as a blueprint for the cell to use when building 
a specific protein.
However, during implementation of this process there can be
many pitfalls. 
\qpar

Consider for instance
the parallel case of the
construction of a building based on an architectural blueprint.
In hot weather concrete
may dry too fast, thus leading to walls with impaired
mechanical characteristics. Freezing weather may 
also cause problems.
In other words, at each step prevailing conditions may not
be optimum or, even worse, may fall outside permissible bounds,
thus leading to structural flaws.
Similarly, inappropriate
temperature or pH conditions may lead to incorrect 
protein production: too little, too much, not at
the right location or not
an appropriate 3-d molecular structure, any of these defects
may possibly lead to life-threatening abnormalities at 
cellular level.
While we are unable to follow such processes
at the level of biochemical reactions, 
visible defects at phenotype level will tell us
that ``something'' went wrong.
\qpar

\qA{What kind of death?}

For lack of a better word, in earlier studies
such events
were referred to as ``production incidents'' or
``manufacturing mishaps'' (Bois et al. 2020). 
In such incidents the blueprint is correct
but not well implemented.
\qpar

Cells and bacteria may experience several kinds
of death.
Our focus is on
the deaths resulting from manufacturing mishaps.
This leaves aside several others forms of death, 
e.g. deaths due
to genetic mutations%
\qfoot{In this respect see Kibota et al. (1996).
The question of the respective weights 
of genetic versus
non-genetic factors was discussed more broadly,
based on observations
of real twins, in the
Appendix of a paper by Bois et al. (2020)}%
, 
senescence deaths due to wear out
and tear 
in growth-arrested bacteria (Yang et al. 2019),
death due to apoptosis,
deaths due to harmful exogenous factors.\qL
Our target could be described by saying that 
the deaths we are interested in occur in the screening process
of new individuals and are deaths for which no other 
identifiable cause
can be found.
\qpar

While most of the paper will focus on how
to count such deaths, we should begin by briefly explain
some of their basic features
(additional explanations
can be found in Bois et al. 2020).

\qA{Motivation}

The first question which comes to mind concerns the
motivation for defining
this new category of defects? 
A very direct answer can be 
found in Fontaine et al. (2008, p.2) 
where it is expressed as follows.
\qdec{``Even in the absence of identifiable exogenous stress,
there remains a measurable, basal death frequency in 
[exponentially] growing
{\it E. coli} populations''.} \qL
Our first purpose is to account for such unexplained deaths.

\qA{Parallels with reliability science}

It is a
fairly natural conjecture to assume that the rules developed
in reliability science,
process management and industrial engineering
also apply here. It would be difficult to illustrate
this assertion through examples of biochemical reactions
at work in the replication process. A typical 
schematic representation
of those complex processes is shown in Fig.2a
from which it is clear that such mechanisms are 
obviously too complicated to
allow a clear insight.
\qpar

Nevertheless a feeling of what is
at stake may be seen by considering
the incubation of an egg. This process
seems much simpler than the replication of bacteria
and yet when one looks closely one realizes that 
there are many requirements.
The creation
of new structures relies on a precise spatio-temporal
organization. Self-assembly processes must take place at multiple
length and time scales. 
\qbu Because an egg shell is a technical
object its manufacturing process allows a fairly clear insight.
Obviously, if too thin or too soft the shell will break
during incubation; on the contrary, if
too hard,
the chick will be unable to break it. In fact 
the egg's characteristics must be consistent
with the strength of the chick. How this
interconnection is implemented, we largely ignore.
\qbu As an example of spatio-temporal organization, 
it can be mentioned that the amount of yolk must
be correlated with the timing of
the chick's development. Too little yolk means
that the chick will starve or will not have
enough energy to break the shell.
\qpar

Although to some readers this case
 may seem simplistic,
it has the great advantage that at each
step one knows exactly what are the optimal
conditions. Moreover, one knows also what will be the effect 
of non-optimal parameters.
Needless to say, for biochemical
replication processes we ignore both the optimal conditions
and the consequences of non-optimal parameters.
\qpar

Neither industrial
production lines nor broody hens can rely on chance. 
From basic ingredients to the final product,
controls
must be set at each step.
Similarly, during the brooding process
temperature and humidity must
remain within narrow bounds, something that can be achieved only
through appropriate controls.
However, the controls 
should not be too numerous or take too much time 
for otherwise the end product will not be ready in time,
e.g. the yolk and albumin will be exhausted
before the growth process is completed. 

\qA{Rules of manufacturing volatility}

Will the introduction of
the notion of manufacturing mishaps give us a better
understanding and even allow us to offer predictions?
We believe so and
the following points give some hints in this respect.

\qbu {\bf Randomness versus deterministic rules}\quad
In the occurence of manufacturing mishaps
there is both a {\it random}
and a {\it deterministic} part. The random part reflects the 
fact, already mentioned,
that usually  the optimal production parameters
are not known. 
The
assumption of randomness (or, if one prefers, stochasticity
that is randomness in time-depended processes) emphasizes 
that, except possibly in special cases, it would
be a hopeless task
to question the exact origin of abnormalities.
\qpar
The deterministic part results from the general
rules of production management already mentioned and
of which illustrations are given below.
\qbu {\bf Why high volatility is to be expected in
long pathways.} The reproduction of a cell
comprises many separate tasks, each of which is a sequential
process involving a succession of steps. \qL
At this point
a remark is in order. There is a common analogy  in which the DNA
is represented by sheet music and the actual cell processes by
the orchestra's performance. One should realize that this 
analogy misses one very important point.
An incident in a performance, e.g. when one of the musicians
momentarily is
out of tune, is of no consequence for the remaining
part of the program. On the contrary, in the construction of 
a building if its foundation or one of the load-bearing walls
are faulty this will jeopardise the whole structure.
As biochemical reactions often work in sequences where the 
products of one serve as reactans for the next,
a flaw at the beginning of the chain will have more serious
consequences than one which occurs toward the end of the chain.
The same argument also suggests that on average
the frequency of mishaps increases with the length
of reaction chains.
\qbu {\bf Greater accuracy requirements make faults more frequent
and more consequential.}\quad As already mentioned,
the production of an egg shell is a fairly critical process,
far more critical than for instance the production of the
albumen content of the egg. In the same line of thought
it can be observed that strabismus is one of the most common 
congenital defects. More generally, 
coordinated movements are a complex processes 
that involve many different muscles,
nerves and parts of the brain.
Any problem in this process may lead to
difficulties%
\qfoot{The medical term for this condition  
is ``developmental coordination 
disorder'' (DCD) or more shortly dyspraxia.
In this framwork strabismus would be called
oculo-motor dyspraxia.}
that will be more consequential (and visible) 
than defects in processes which do not require the
same level of coordination. In short, the higher the complexity
of a manufacturing process, the higher the risk
of error propagation.
\qbu {\bf Is a faster division time an aggravating factor?}\quad
In industrial production
when the time allocated to a given task is reduced
less controls are performed which
in turn may result in more defects going undetected.
Something similar can be conjectured in
the process of bacterial replication. To put numbers
on this issue let us mention two populations
studied by Kelly and Rahn (1932). They observed
divisions of {\it Bacterium aerogenes} whose 
average division
time was 30mn and buddings of {\it Saccharomyces 
ellipsoideus} whose average budding times was 105mn,
i.e. 3.5 times longer.
If our conjecture holds the fastest process
should have a higher death rate. 
\qbu {\bf Are volatility and deaths correlated?}

As a characteristic 
easier to identify than death, one is tempted
to consider volatility%
\qfoot{The term ``volatility'' is understood here with the same
meaning that it has in finance, i.e. it includes all statistical
estimates of variability.}
in daughter's features. After all,
occurrences of deaths may be seen as cases of 
volatility becoming so excessive that stretched features
become incompatible with the continuation of life.
\qpar

It is well known that at the beginning of the exponential  phase
the reproduction process is so fast that replication starts even before
separation is completed. If our previous conjecture 
is correct, one should  see in this phase a volatility that is 
higher than in the subsequent, slower part of the exponential
growth.

\qA{Perspectives opened by the investigation of congenital defects}

The previous points can be summarized by saying
that any information about a process (e.g. length of passways,
accuracy requirements) will allow us to
propose conjectures and predictions concerning congenital
anomalies. 
\qpar

Moreover, observation of manufacturing mishaps 
in the exponential
growth phase is the simplest case one can think of. 
There are two reasons. \qL
(i) The fact that all cells are clones
limits genetic effects.
\qL
(ii) It has
been emphasized (see Sezonov et al. 2007) that during the 
exponential phase all intrinsic parameters of the cells
(e.g. their macromolecular composition) remain
constant. \qL
In other words, through its simplicity, this
system should provide a valuable opening, just as the study of
the hydrogen atom opened the way to
atomic spectroscopy.
\qpar

The rest of the paper proceeds as follows.
\qbu In section 2 we draw on the mechanisms described
above to predict possible regularities 
to be observed in infant mortality rates.
\qbu Section 3 gives accounts of some former attempts
at measuring post-fission death rates. It appears that
this issue already attracted attention
a long time ago. 
\qbu In the penultimate section prior to the conclusion we suggest
a possible protocol based on modern techniques.

\qI{Expected regularities of infant mortality rates}

First, we wish to explain the exact meaning given here to the
expression ``infant mortality''. 

\qA{Meaning of infant mortality}

In medical terminology ``infant mortality'' means mortality
during the first year after birth. Conceptually this definition
is not satisfactory however. 

%
\begin{figure}[htb]
\centerline{\psfig{width=17cm,figure=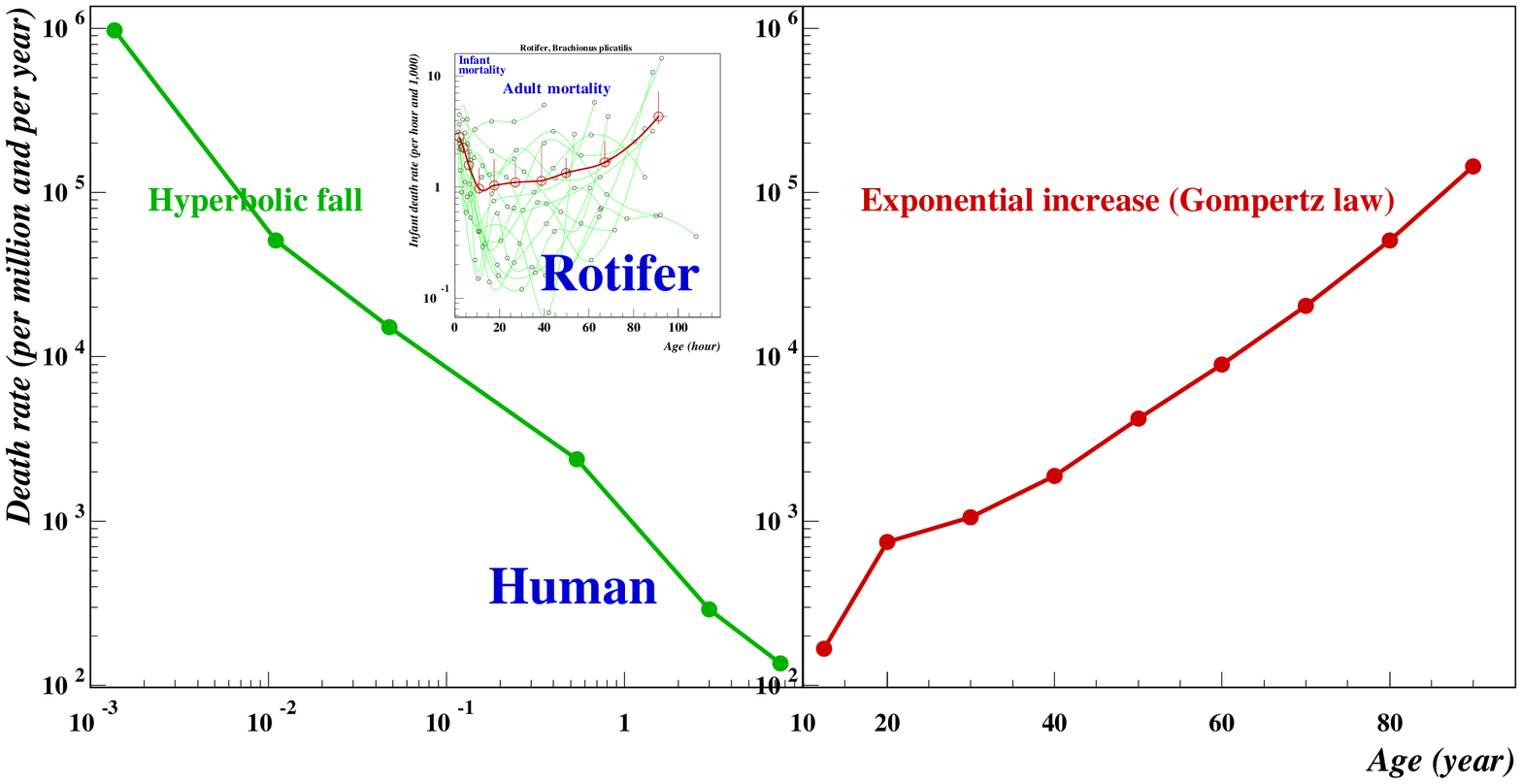}}
\qleg{Fig.1a Infant and adult mortality in humans
and rotifers.}
{For human newborns very accurate mortality data
are available 
which extend from a few hours after birth to late childhood.
In this phase the infant mortality decreases as a power law
of the form: $ y=1/x^{\alpha} $, where the exponent $ \alpha $
is close to 1. Whereas for humans the phase of infant mortality
lasts about 10 years for rotifers (swimming animals of a size
of about 50 micrometers) it lasts only 10 hours.}
{Sources: The statistical data are for the United States
for the period 1999-2016 and are taken from the CDC
data base for detailed mortality. The rotifer data are
from an experiment conducted in 2019 and described in
Bois et al. (2019).}
\end{figure}

%
\begin{figure}[htb]
\centerline{\psfig{width=12cm,figure=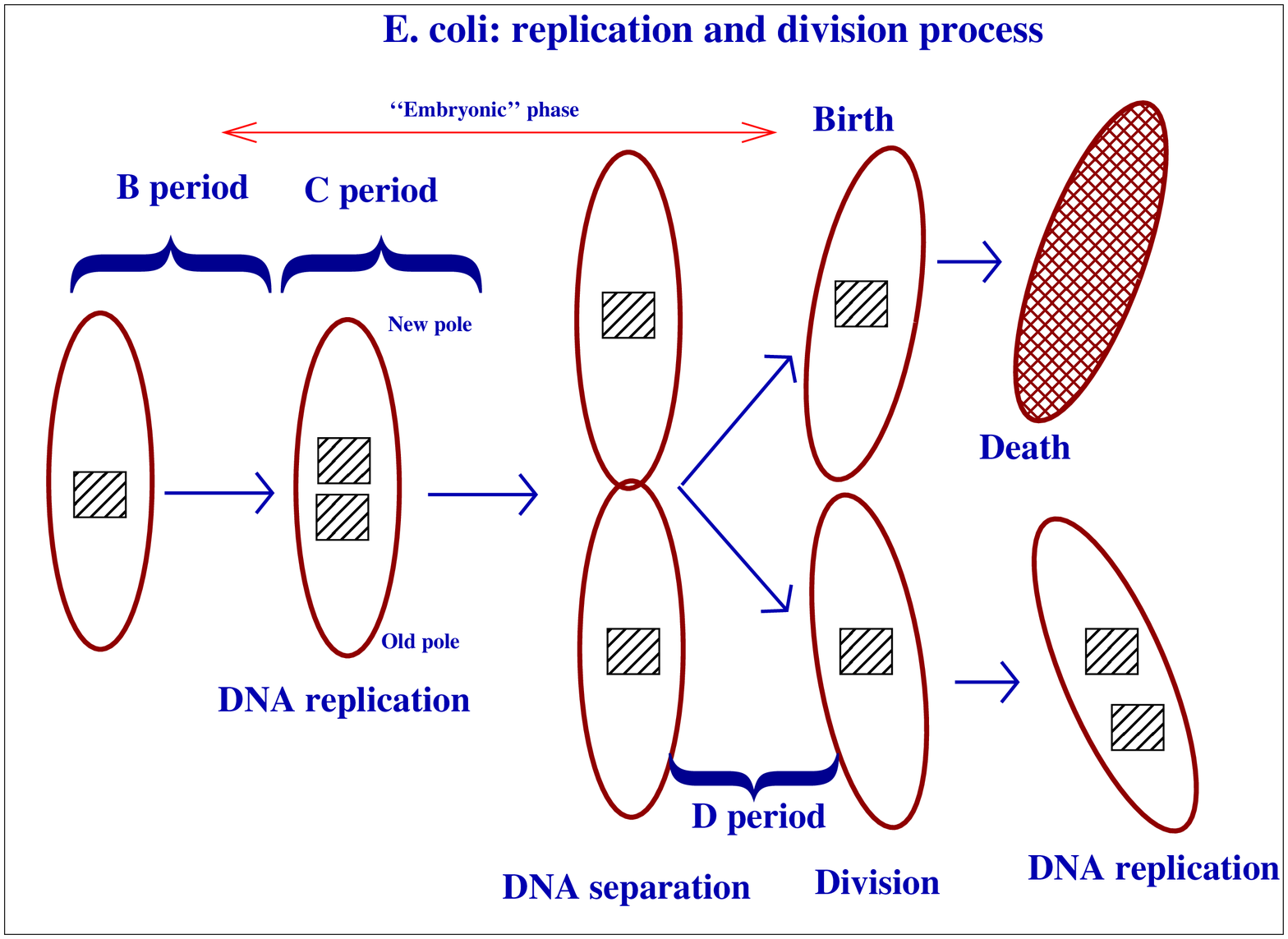}}
\qleg{Fig.1b Schematic representation of the replication
and division process of E. coli.}
{At first sight reproduction by fission may seem very different
from reproduction in multicellular organisms. However, some parallels 
can be drawn. 
The phase between replication of the DNA and separation
of the two cells can be seen as the combination of two processes.
First, the production of an oocyte
followed by a process of embryogenesis and then finally the 
transformation of the embryo into a newborn by hatching of eggs
or otherwise. Needless to say,
this is a very schematic representation. For instance the DNA is
shown concentrated whereas it is rather spread throughout
the wole cytoplasm. Fig.2a shows again these steps 
(though even more schematically) because they are the 
starting point of our experiment. The periods B,C and D 
are a standard periodisation of the whole reproduction
process.}
{}
\end{figure}

What distinguishes newborn mortality
from adult mortality is the fact that the first 
{\it decreases} with age
whereas the second {\it increases} with age in conformity with
Gompertz's law (see Fig.1a). 
However, the decrease phase does not end
after the first year, it continues until through the age of 10.
A coherent definition of ``infant mortality'' should
include the whole age interval during which the mortality rate
decreases. \qL
Actually, the fact that after the age of 10 the mortality rate
starts to increase does not mean that the deaths
due to congenital factors come to an end, it rather means
that adult mortality becomes dominant. 
Seemingly minor congenital defects can become a cause of 
death much later in life.
We return to this point later.

\qA{Birth ends ``compensation effects''}

What is meant by the expression ``compensation effect''?\qL
In a multicellular organism each organ involves many cells,
as a result the death of one (or even a small number)
of them will have no serious consequences because the remaining cells 
will try to compensate the lost capacity. When there are only two
cells, a shortage in one may still be compensated
(at least partially) by the other. This effect is illustrated
in Fig.2a.  A shortage of, say protein $ A $ is compensated
through transfer from the other cell.
However, separation ends
such compensation mechanisms. After birth, human
newborns must be able to breathe; similarly after division,
the daughter cell must be able to produce all 
proteins it needs.

\qA{Accumulation of defects}

The accumulation of defects is a phenomenon which may be seen as
specific to 
organisms which reproduce by division. The reason derives
immediately from the argument just given.
If one supposes that the shortage of $ A $
is not lethal and does not prevent reproduction, this
defect will remain in existence on the mother side
in subsequent generations. 
Such a scenario is consistent with the phenomenon
of aging demonstrated in Stewart et al. (2005).
\qpar

Although somewhat different in
its occurrence, the phenomenon of defect accumulation is akin to
the gradual wear-out that occurs in the aging process 
of multicellular organisms. As an illustration, consider
osteoporosis; although it becomes visible only in old age,
in fact it starts around the age of 25.
There are many other examples, e.g.
the ability to hear high frequencies or maximum heart rate 
capability, or progression of stiffness in heart valves.
What distinguishes a heart valve problem in old age from a
neonatal heart valve problem is not the organ involved which
is the same but the fact that the first comes gradually
whereas the second is triggered by a massive defect
and leads to death within a few days.
\qpar

One of the most fragile components of bacteria seems to be
their membrane. In this case it is easy to imagine 
the same two regimes:
(i) a gradual erosion of the membrane due to growing inability
to produce the required proteins versus (ii) a dramatic failure 
in the form of a breach due to a lethal manufacturing defect.
If massive enough, it could kill both daughter
and  mother cells if it occurs
during the short time interval between ``production''
of the daughter cell (by its own DNA) and separation.
of the two cells.
\qpar

The fact that membrane failure is a major cause of death
is demonstrated by the data of Table 1, namely the
near equivalence of deaths identified
by stains (i.e. resulting from breached membranes) 
and the deaths identified by immobility and inability to
divide further.

%
\begin{figure}[htb]
\centerline{\psfig{width=16cm,figure=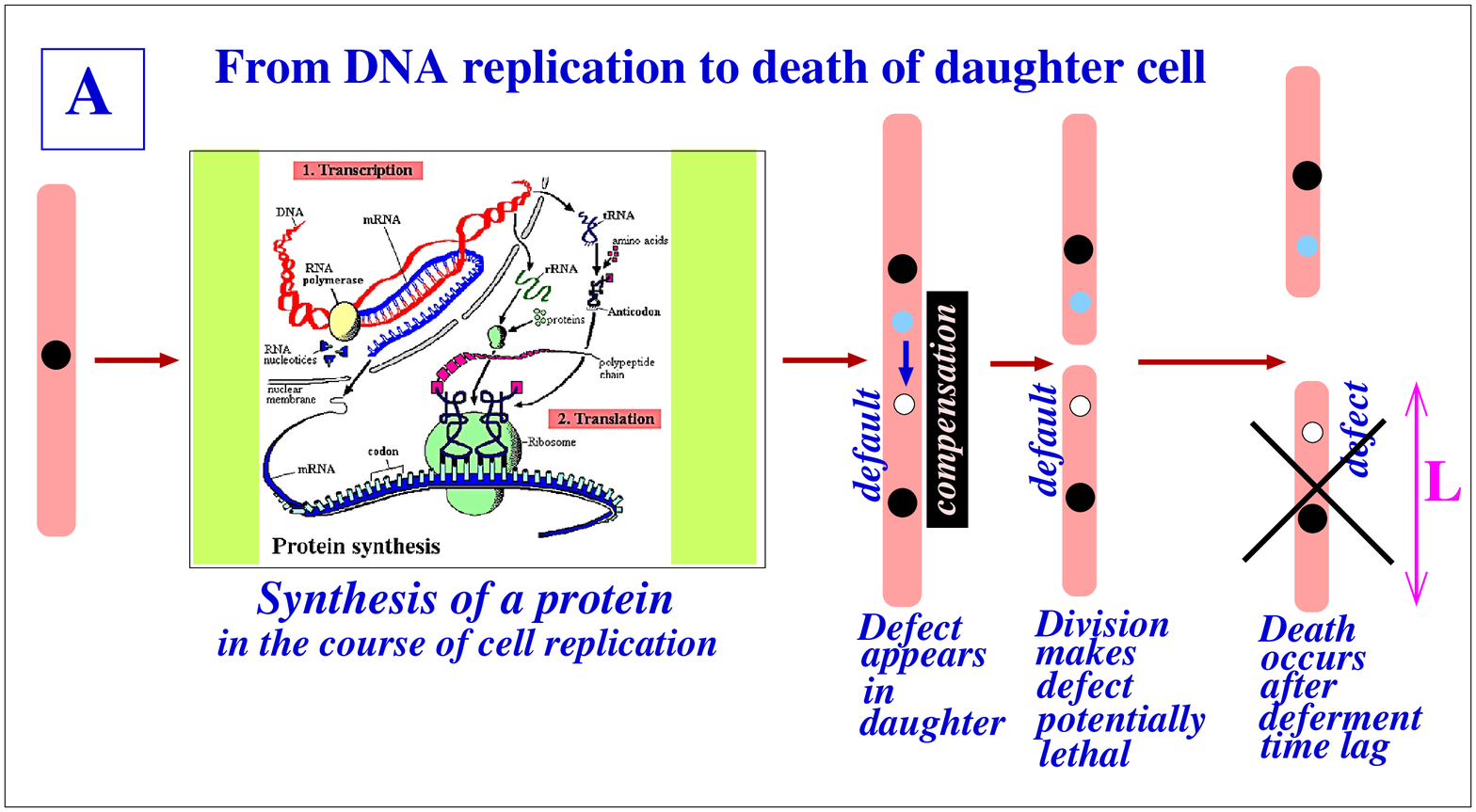}}
\qleg{Fig.2a Occurence of a defect in a daughter cell.}
{The main purpose of the picture on the left hand side
is to suggest that the
synthesis of a single protein macromolecule is a process
which involves many successive steps each of which
comprises a chain of appropriate biochemical reactions. 
Reactions occuring in non-optimal conditions 
(e.g. inappropriate temperature or pH level) may
affect the structure of proteins, enzymes and other
important molecules. Eventually, this may lead to a default
(schematized by an empty circle). At first the default may
not be of great consequence for
as long as the two cells are tied together a default 
in one can be compensated by the other. After division
each cell must be able to live on its own resources
which is likely to make the consequences of any defect much 
more serious (there is a similar situation for newborns 
immediately after birth). It is for the purpose of clarity
that the nucleus are shown as black cercles.}
{Source for protein synthesis picture: Slideshare website 
(public resource).}
\end{figure}

The process leading to a cell death is schematized in
Fig.2a. Incidentally, 
human infant mortality strongly
suggests that non-lethal defects largely outnumber lethal
defects. However, for bacteria it is very dificult to
identify non-lethal defects. The death rate gives
only a crude estimate of the most
serious manufacturing defects but
at least it can be measured.
\qpar

One obstacle to sound measurement is of course the well-known 
inherent variability (of the order of 30\%)
across generations
in the lengths of {\it E. coli} at same
stage of their life cycle (for instance at birth). 
As this variability seems to be truly random (see. 
Adiciptaningrum 2015) it is possible to get rid of it
through the time-honored method of taking averages over a large
number of repeated events.

\qA{How can ages at death be estimated}

Once the magnitude of the death effect has been asserted
the second objective is to estimate the age at death.
As in {\it E. coli} age is strongly 
correlated with size, one can use the latter to get age
estimates. For this purpose rod-shaped organisms
are particularly convenient.

\qA{Conjecture for the post-birth death rate}

In all species studied in
Bois et al. (2019) the post-birth age-specific death rate
was found to be a decreasing function of age.
More precisely, it was found to be an hyperbolic fall.
We conjecture that it will be the same here.
\qpar

The experiment
which should allow us to observe the age-specific death rate
and check this conjecture is schematically
summarized in Fig.2b,c. Let us explain how the
hyperbolic fall conjecture can be tested.

%
\begin{figure}[htb]
\centerline{\psfig{width=16cm,figure=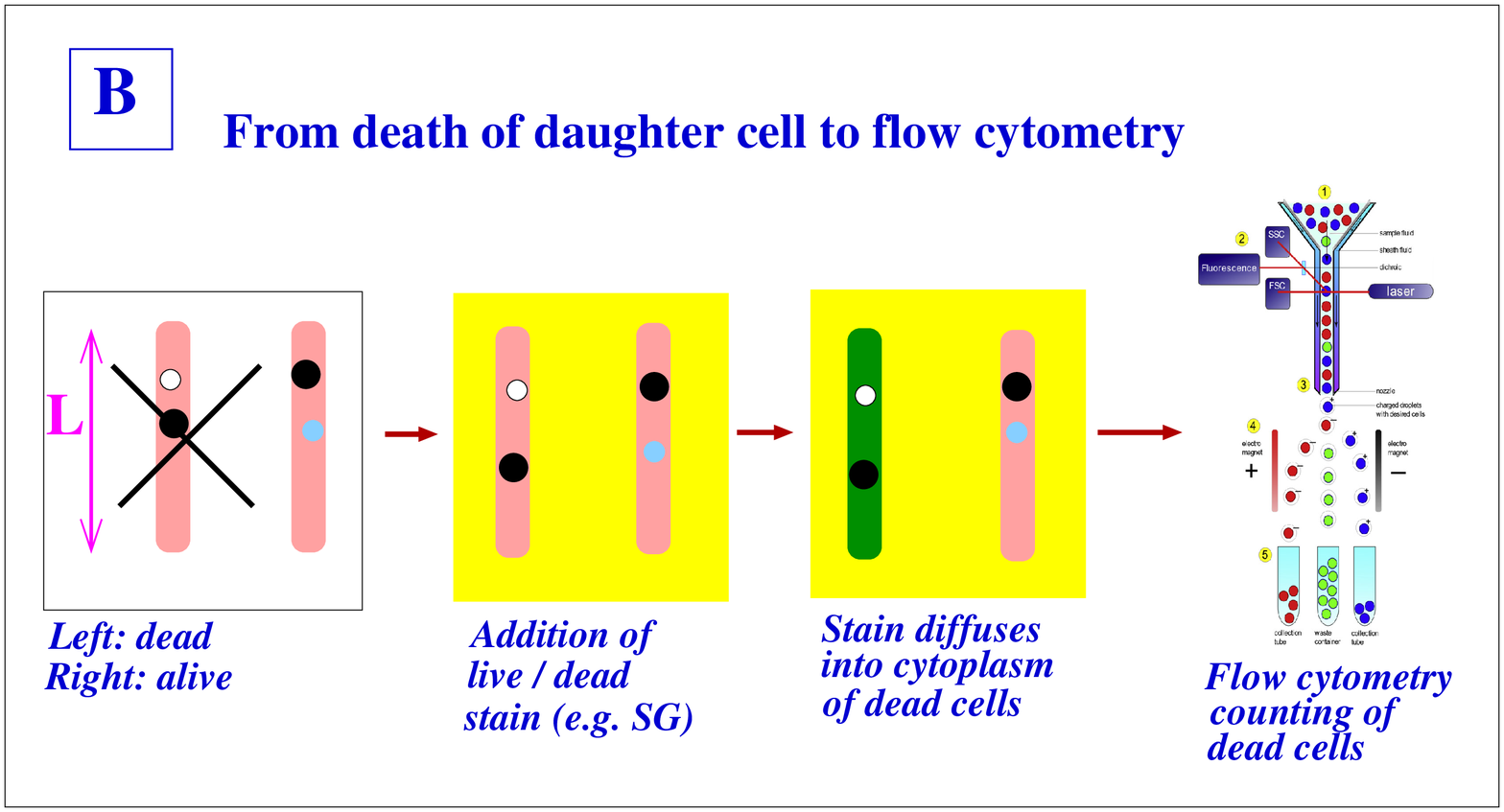}}
\qleg{Fig.2b: Identification of dead cells by life / death
stain.} 
{Staining and cell counting by flow cytometry (FC)
work hand in hand. As the stain can diffuse into
the cell only when the cell membrane is breached,
the measurement is likely to provide a lower bound 
of the death rate. A clear advantage of the FC technique
consists in the size of the samples that can be treated.
Whereas in the observation of
invidual cells the sample-size in each assay
is limited to a few hundreds, in FC samples of one million
can be tested in each run. SG which means Sytox Green
is the commercial name of a life/death stain. It can diffuse
into a cell whose membrane is breached.}
{Source: The FC picture is from Internet (public resource)}
\end{figure}

%
\begin{figure}[htb]
\centerline{\psfig{width=16cm,figure=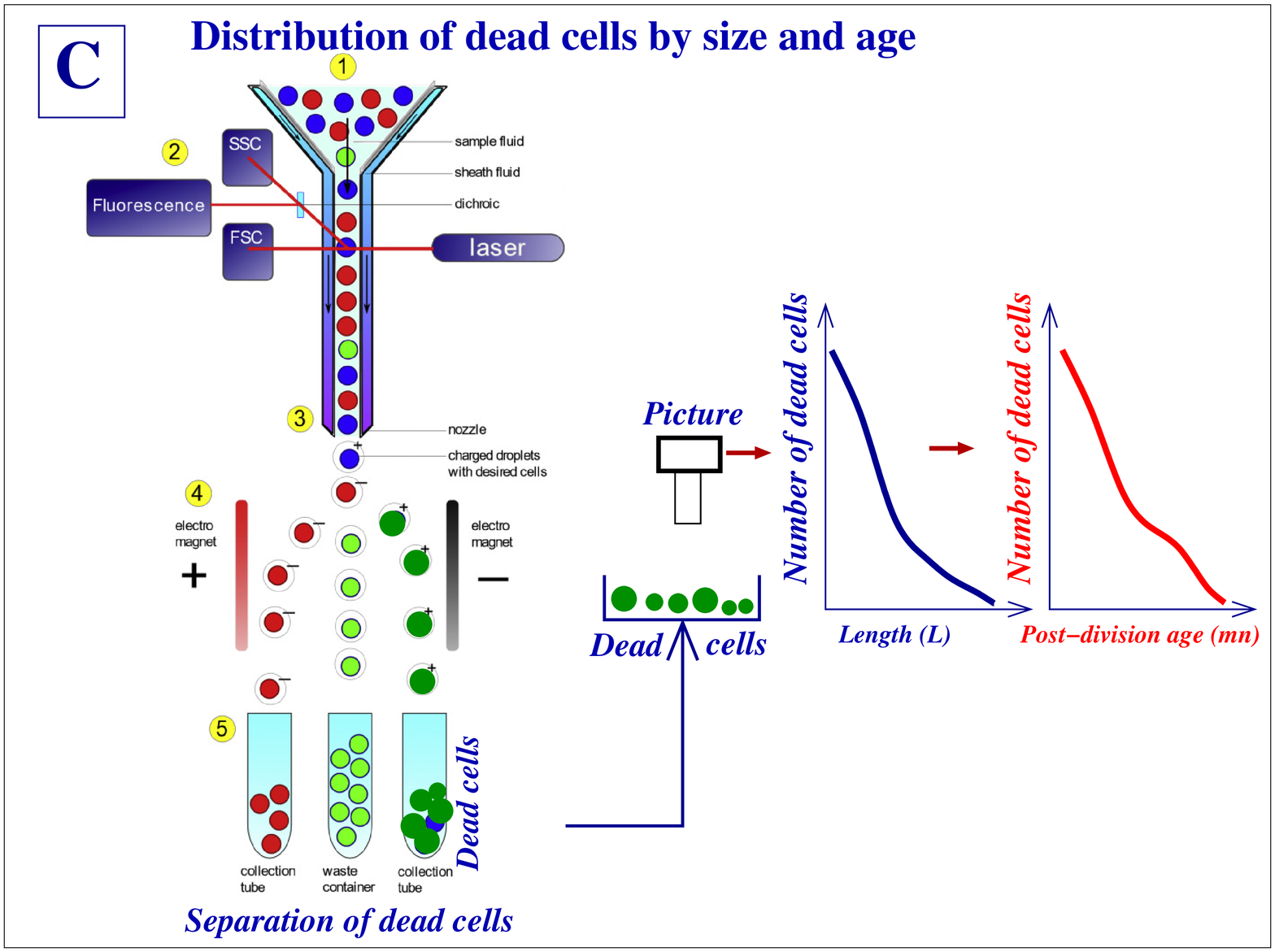}}
\qleg{Fig.2c: Distribution of dead cells by age.}
{The age-specific infant mortality has a characteristic
shape (explained in the text). The droplets on the right-hand 
side contain the dead cells which will be investigated 
separately. The two other droplet releases contain the living cells
and may be cells selected through a third criterion.
Incidentally, 
most flow cytometers have the ability to measure
the size of the cells at the same time as they 
identify the dead cells, but one would expect this capability
to be more reliable for spherical bacteria than for rod-shaped
bacteria. The later allow a better accuracy for the simple
reason that for a given change of volume $ \Delta V $ 
the change of the length is: $ \Delta L \sim \Delta V $, whereas
the change of the radius of a sphere 
is only: $ \Delta R \sim [\Delta V]^{1/3} $. 
In the figure the acronyms FSC and SSC mean forward scattering}
and backward scattering respectively. The expression ``stealth 
fluid'' refers to a layer of fluid which ensures that the flow 
is laminar. This is particularly important for 
reliable measurement of the length of rod-shaped cells. 
The  end step
microscopic examination of the dead cells has a two-fold
purpose:
(i) to control the reliability of the length estimates 
(ii) to allow a close examination of the dammaged cell membranes.
{}
\end{figure}

Assume that in 100à,000 daughter
cells which emerge from the fission process
there are some 100 with
potentially lethal defects that can be graded from
very severe to somewhat less severe (the analog in human newborns
would be heart defects ranging from misshapen hearts
to only malformations of the aortic valve). 
The cells with first grade defects 
will die immediately, those with grade 2 severity
somewhat later, and so on. As more and more
cells die, those which remain will experience a smaller
death rate. 
\qpar

Let us divide the age interval following fission 
into subintervals in the following way (time units are minutes):
$$ I_1=(1,2), I_2=(2,4), I_3=(4,8), I_4=(8,16) $$

It was found in Bois et al. (2019) that the numbers
of deaths in each of these subintervals are nearly
the same. This rule relies on an age-specific death rate curve
of the form: $ y=C/x^{\alpha} $ where $ x $ is age, $ y $ the
number of deaths and $ \alpha $ an exponent that is close to 1.  
This rule means that, in infant mortality, 
if age is multiplied by 2, $ y $ is divided
by 2. \qL
If, as suggested earlier, age is estimated by the length
of the bacteria, the previous statement can be
easily reformulated in terms of size.
If the pattern described above can be observed it would 
be a further confirmation of a regularity already observed
in other organisms.

\qI{Toward accurate measurements of post-fission death rates}

Attempts to measure the death rate of individual bacterial cells
go back at least to the beginning of the 20th century.
Rather than to provide a historical review, our goal is to show
that
this problem has been recognized as important for a long time.

\qA{Paper of Kelly and Rahn (1932)}

This is by  no means the first paper on this problem
but it  provides the clearest statement of why
it is a key-issue. The first lines of the paper are worth being
quoted verbatim.
\qdec{``It has been assumed by many bacteriologists that
during the period of rapid growth, in a satisfactory 
culture medium, some bacteria {\it will die} in spite of
good food and favorable environment. No doubt this assumption 
was derived from an analogy with populations of higher
forms of life, of which a number of individuals are
known to die before they reach the reproductive age even with
good care.'' (Kelly et al. 1932, p.147)}
\qpar

Using a method pioneered by J. Orskov (1922),
the authors followed the replication of individual bacteria 
(and one yeast species)
on a solid medium. They recorded the family trees spanning 
4 generations. Altogether they observed 1,766 divisions.
977 of {\it Bacterium aerogenes}, 325 of {\it Bacillus cereus}, 
464 {\it Saccharomyces ellipsoideus}.  On average the time intervals 
between fissions of  {\it Bacterium aerogenes} was 30 mn but with great
inter-individual fluctuations (coefficient of variation 
$ \sigma/m $ close to 100\%). 
\qpar

The observations led the authors to the following conclusions.
\qbu There was not a single instance where a cell ceased 
multiplying or became dormant.
\qbu In the very words used by the author, there was not
a single case of ``infant mortality'', that is to say 
a cell which died after division. 
\qpar

The first conclusion is reassuring because, even today, 
researchers often worry about the risk of mistaking
dormant cells for dead cells (see the discussion
at the begining of Garvey et al. 2007).
\qpar

The second conclusion is also interesting for it shows
that if such an infant mortality exists (as is indeed
observed in experiments done in the
past two decades as reported below)
then its rate is lower than: 1/1766 = 0.56 per 1,000.
(for a time interval of 30mn). This is a rough estimate
obtained by bulking together all three species. 
Estimates detailed by species (at least for the two largest)
are given in Table 1.

\qA{The Wilson paper (1922)}

Ten years before the article by Kelly and Rahn, Wilson was
already calling into question the dogma that in a young broth
culture (up to 24 hours)
all bacteria are living. Here is what he writes at the beginning
of his paper.
\qdec{``On looking up the literature it was found that of 
the many observers [the author cites 9 papers published
between 1898 and 1920]
who had made a comparison of the two counts
[namely the total number of cells 
that can be counted in a culture, whether dead or alive
on the one hand
and the number of viable cells defined as cells that 
are able to fission on the other hand], 
the discrepancy was passed over with
little comment.
As in this stage all bacteria were assumed alive
any discrepancy had to be due to errors.''}
\qpar

Then, in his lengthy 41-page long paper the author 
examines one by one
all successive operations required in a counting procedure.
He tries to make them as rigorous as possible and he estimates
the remaining error margin. 
For instance dilution before counting is an operation
which requires great care but nevertheless cannot be made very
precise. In fact, the author was facing an impossible task. 
It should be realized that despite being a standard counting 
technique in the 1920s, this
was completely different from the observation of individual
cells as pioneered by
Kelly and Rahn (1932). Such global counting techniques were beset
by too many uncertainties.
Therefore, it is not surprising that Wilson arrives at vague and
disappointing conclusions (p.444). 
\qdec{``It seems that in cultures of {\it Bact. suipestifer}
there is a normal death rate even during the period of maximum
growth. Its extend will vary from culture to culture.
In some it is as high as 43\%, in others it is only 20\% or 10\%,
while finally in a few it is for a short period actually nil.}
\qpar

The Wilson paper makes us realize that the observation of
individual cells pioneered by Kelly and Rahn represented
a breakthrough. It is true that they
were unable to see any death in
the exponential phase but that was only because their
sample of 1,766 cells
was too small. Their methodology was sound and
opened the door to further observations, possibly
with larger samples. Yet, this
did not happen until 73 years later. Indeed, in 2005 this
investigation was
resumed.
Thanks to modern computerized counting techniques
and a sample some 20 times larger 
namely 35,000 {\it E. coli} cells,
it revealed some 16 deaths; see Stewart et al. 2005%
\qfoot{In fact, this was not the only objective of the
investigation. Another purpose was to study
the erosion of the fission ability
in the course of successive generations. 
Such an erosion was seen as an indicator of aging.}%
.
\qpar

Whereas the measurement of Stewart et al. (2005) relies
on the observation of individual cells, a paper of 2008
by Fontaine et al. relies on a global (not individual)
observation.
It is to the discussion of these modern investigations
that we turn now.

\qA{The paper by Stewart et al. (2005)}

In the Kelly and Rahn (1932) experiment
successive divisions were followed over
4 generations, a process which from each single initial cell
produced $ 2^4=16 $ cells.
In the Stewart et al (2005) experiment (subsequently ``Stewart 05'')
up to 9 generations were followed, a process through which each
initial cell gave rise to $ 2^9 =512 $ {\it E. coli} cells.
Time-lapse images were taken and analyzed automatically thanks
to a dedicated software. As 94 colonies were
analyzed this led to a total of 35,049 divisions. 
\qpar

The criterion used for the definition of death was immobility
combined with no growth. Some 16 cell death were observed.
Unfortunately their sizes were not included in the publication
because the main purpose of the paper was in fact the detailed
study of aging.
%
\begin{table}
\centerline{Table 1: Bacterial death rates in the 
exponential phase}
\small
\vskip 2mm
\hrule
\vskip 1.5mm
\hrule
\vskip 2mm
$$\matrix{
 &\hbox{Year} & \hbox{Paper} \hfill & \hbox{Method} \hfill & 
\hbox{Sample} & \hbox{Number} & \hbox{Death rate (dr)} \cr 
 & & \hbox{} & \hbox{} & 
 \hbox{size} & \hbox{of deaths} &\hbox{per 1,000} \cr
\qtb
 & & \hbox{} & \hbox{} & 
 \hbox{} & \hbox{} &\hbox{and per hour} \cr
\noalign{\hrule}
\qth
1& 1922 & \hbox{Wilson} & \hbox{Global} & 
\hbox{undefined}  & \hbox{unreliable} &\hbox{unreliable} \cr
2& 1932 & \hbox{Kelly (1)} & \hbox{Individual} & 
  733 & <1 & \hbox{dr}<4.1 \cr
3& 1932 & \hbox{Kelly (2)} & \hbox{Individual} & 
  420 & <1 & \hbox{dr}<1.4 \cr
4& 2005 & \hbox{Stewart} & \hbox{Individual} & 
  35,049 & 16 & 1.5 \cr
\qtb
5& 2008 & \hbox{Fontaine} & \hbox{Global} & 
 10^6  & 700 & 0.7 \cr
\noalign{\hrule}
} $$
Notes: ``Global'' means measurement performed on a 
large number of cells in liquid medium.
For this global measurement
it is the technique of flow cytometry which brought
about a breakthrough and allowed reliable measurements.
The following bacteria and yeast were investigated (in parenthesis
is the length in minutes of the reproduction cycle).
1: {\it Bact. suipestifer} and other species, 
2: {\it Bacterium aerogenes} (30mn),
3: {\it Saccharomyces ellipsoideus} (105mn), 
4: {\it E. coli} (30mn), 5: {\it E. coli} (30mn). For case 5
it was assumed that the state having an optical density
$ \hbox{OD}_{600} $ around 0.2 lasted one hour.
\qL
Sources: Based on the papers cited in the third column.
\vskip 2mm
\hrule
\vskip 1.5mm
\hrule
\vskip 2mm
\end{table}

\qA{Paper of 2008 by Fontaine et al. (FC measurement)}

Beyond its specific purpose, this paper 
(thereafter ``Fontaine 08'') can be seen
as a continuation of the previous one as is 
indeed confirmed by the participation of two 
authors in both papers. 
\qpar

In the technique that is used here
the cells are not monitored individually.
Instead, the recourse to flow cytometry allows global estimates.
Stained dead cells, 
or more precisely those cells whose breached membranes
allow the stain to drift into the cytoplasm,
are counted thanks to a flow cytometer (FC).
In such a device the light of a laser is diffused by the
cells when they move through the beam, then received by a sensor
and amplified by a photomultiplicator and finally analyzed by
a computer software algorithm.
Flow cytometry allows many characteristics of
the cells to be identified and
recorded. Here this technique is used to
count stained or fluorescent dead cells. 
\qpar

Flow cytometry
began to be used in the 1950s and was really a game changer.
It replaced the successive manipulations 
that we mentioned in our account of the Wilson paper of 1922.
Whereas
improving their accuracy was an impossible task,
FC provided a completely new approach which proved
effective and reliable.
\qpar

Although FC is the key of the measurement method,
a number of additional verification tests are required
to ensure that what is measured by the device is indeed
the appropriate death rate.

\qA{Accumulation and amplification of mishaps}

Here we come back to the question of the accumulation
of defects in order to see how the shape of the age-specific
death rate may be affected.
\qpar

Assume that in generation 
$ k $ a protein (call it $ P $) has been produced in
insufficient quantity, then in the next 
replication process there are two possibilities. 
\qbu The problem of the low level of $ P $ is identified
and corrected by an appropriate temporary overproduction.
Under this assumption generation $ k+1 $ will have regained
its nominale characteristics.
\qbu Here we assume that
the problem, either is not identified or, if identified,
cannot be corrected because under current conditions
increased production of $ P $ is not possible.\qL
In this case the default will not be corrected and may even be
aggravated. Even though the initial mishap was not lethal
it may become so in generation $ k+1,\ k+2 $ or later on.
\qpar

Note that in generation $ k+1 $ the default is likely to be 
shared equally by  the mother and daughter which means that the 
compensation mechanism illustrated in Fig.2b cannot take place.
Therefore, the separation of mother and daughter will
not have the same filtering effect. Under this assumption
there is no reason for death to occur specifically
in the first moments after separation. This phenomenon would 
give a fairly uniform distribution of deaths with
respect to age.
\qpar

In short, according to the previous discussion, a fairly
uniform distribution of cell deaths over their life time
would suggest a mechanism of gradual amplification 
of mishaps. The occurence of pair-wise deaths (meaning
deaths of both mother and daughter at about the same age .
after division) would  be an indication pointing in the
same direction.


\count101=0  \ifnum\count101=1

\qA{Discussion of earlier papers on error propagation}

In the 1960s and 1970S
there have been a few studies about errors in the biosynthesis 
of proteins, e.g. Loftfield 1963 and Loftfield et al. 1972%
\qfoot{On p.1353 the authors say that the paper of 1963
was the ``only reported attempt to determine chemically the
frequency of translation errors''.}%
, 
Gallant 1979.
However, only errors of a few specific types were considered
e.g. mistranslations from messengerRNA. Such observations 
led to estimates of error frequency of 3 parts in 10,000
(Loftfield et al. 1972, p.1356). 
\qpar

It can be observed that the
word ``error'' refers to dichotomic mishaps of ``yes - no'' type.
Most likely one should also consider quantitative mishaps
of the ``too little - too much'' type as considered in the
previous subsections.

\fi

\qA{Summary of death rates measured in the exponential phase}

Table 1 presents the death rates results found in the papers
discussed. They cluster reasonably well around an average
value of  1 per 1,000 and per hour.
\qpar

For red blood cells, using data found on the Internet
(30 trillion red blood cells in total, 2 million die per second)
one gets a death rate of: 0.2 per 1,000 and per hour.
These cells are somewhat special in the sense that they
have no nucleus and therefore cannot make proteins to repair
themselves. Their life span is known to be of the
order of 120 days.
Thus, one would expect their death rate
to be a kind of upper bound.

\qA{Life-death separation based on buoyant density}

The experiment described in Fig.2a,b,c comprises two
successive steps, first the cytometric life-deaf separation
and secondly the  optical determination of
the distribution of the sizes of the dead cells.
\qpar

Instead of the cytometric separation, can one 
imagine a separation based on buoyant density?
The idea is not altogether absurd. Several
studies (Pierucci 1979 Table 1, Woldrigh et al. 1981 Table 1)
have shown that the specific gravity of {\it E. coli} cells is
about 1.1. Moreover, whereas the length of the cells is
affected by the growth rate (the variabilty reaches 40\%), 
the specific gravity is very stable (variability 
lower than 2\%.. This means that in water%
\qfoot{The specific gravity of an LB medium is almost 1.}
 dead cells will sink to the
bottom whereas, due to their mobility, living cells will
remain distributed between bottom and surface. The key-question
is how fast they will sink; it can be answered based on
Stokes law. This law applies whenever the Reynolds number
is smaller than 1 which is obviously true for objects
as small as bacteria. A falling cell experiences 
three forces: (i) downward gravity force  and two upward forces:
(ii) buoyancy (iii) drag given by Stokes formula. Once the
velocity has become stationary the sum of the three forces
is zero and from this equation one can derive the stationary
velocity $ v $ in the following form.
$$ v={ d^2 g (\rho_c - \rho_w ) \over 18\mu} $$

where:\qL
$ d $: Equivalent spherical diameter of a cell rod; about 1 cubic
micrometer. \qL
$ g $: acceleration of gravity, $ g=9.81 m/s^2 $ \qL
$ \rho_c,\ \rho_w $: specific gravityy of cell and water respectively,
$ \rho_c =1.1,\ \rho_w=1.0 $ \qL
$ \mu $: Dynamic viscocity of water. $ \mu=8.9\times 10^{-4} $
Pa$ \times $s. \qL
One gets: $ v=6 $ micrometer/minute, a speed which is of course
too slow to be useful. In the present case one cannot use
centrifugation for the centrifuge force would overpower the tiny
upward sustaining force generated by the cells. The velocity
$ v $ may be substantially increased by replacing water by
a liquid of lower specific gravity and lower viscosity.
\qpar

The previous calculation can tell us something else 
which may be of interest for the step following cytometry.
When one 
uses centrifugation 
to increase the concentration in dead cells of the solution provided
by the cytometer, the previous result shows that to get
a velocity of 6mm/mn one must
replace $ g $ by $ 1000g $. As, however, the shape and size of the
cells may be affected by such a high acceleration, we see that 
it is certainly better
to use only $ 10g $, yet applied during 100 mn. in order to get the
same separation.

\qI{Conclusions}

The main incentive for measuring age-specific death rates
is to verify whether our understanding of infant mortality
gained from the observation of multicellular organisms
is correct. The conjectures proposed in the present note will
offer instructive tests.
\qpar

Human infant mortality helps us to understand congenital
anomalies. In a similar way, the infant mortality of microorganisms
reflects the mechanisms which rule their life cycle.
Today new technologies provide
a novel route to test our conjectures and give us a better
understanding of questions which have challenged scientists
for well over a century. Preliminary cytometric measurements
(which require confirmation)
suggest the existence of a mortality peak in the minutes 
prior to separation.
\qparr

We have summarized the evidence already available and we have
outlined a protocol in the hope that it will give our
readers an incentive to carry out the experiment.

\vskip 3mm

{\bf Acknowledgments} \quad We are indebted to many persons 
for their help and advice.
Since right after its inciption in late
2019 (Bois et al. 2019, p.14),
this research project benefited from Dr. Arnaud Chastanet's interest
and expertise. The visits of one of us (BMR) to his INRAE (National
Institute for Agricultural and Environmental Research) laboratory
were important steps  We were initiated to cytometric analysis
by Annie Munier and Angelique Vinit; many thanks to them for 
their expertise and interest in our project.
Thus, little by little, 
thanks also to the help of Dr. Guennadi Sezonov and
Dr. Ivan Matic, confidence developed gradually and 
what was initially just a conjecture became an experiment which
may now be carried out with some chance of success.

\vskip 3mm

{\bf Ethical statement}
\qee{1} The authors did not receive any funding.
\qee{2} The authors do not have any conflict of interest.
\qee{3} The study does not involve any
experiment with animals that would require ethical approval.
\qee{4} The study does not involve any participants that
would have to give their informed consent.

\vskip 7mm

{\bf References}

\qparr
Adiciptaningrum (A.), Osella (M.), Moolman (C.), 
Lagomarsino (M.C.), Tans (S.J.) 2015:
Stochasticity and homeostasis in the 
{\it E. coli} replication and division.
Scientific Reports (Dec 2015) \qL
[The paper reports that the variability of the lengths
of {\it E coli} cells at the same stages of their life cycle
is of the order of 30\% even for
cells with identical genome under stable conditions.]

\qparr
Bois (A.), Garcia-Roger (E.G.),
Hong (E.), Hutzler (S.), Irannezhad (A.), Mannioui (A.),
Richmond (P.), Roehner (B.M.), Tronche (S.) 2019:
Infant mortality across species. A global probe of congenital
abnormalities.
Physica A 535,122308 (Oct. 2019).

\qparr
Bois (A.), Garcia-Roger (E.G.),
Hong (E.), Hutzler (S.), Irannezhad (A.), Mannioui (A.),
Richmond (P.), Roehner (B.M.), Tronche (S.) 2020:
Congenital anomalies from a physics perspective.
The key role of ``manufacturing'' volatility.
Physica A 537,122742.\qL
[Although published after the previous one, in logical order
this paper should be read first. 
It emphasizes that, in a general way,
besides genetic mutations and environmental
factors, manufacturing defects are an important cause
of congenital anomalies and infant mortality.
This effect leads to the prediction
of death spikes at the beginning of embryogenesis
and at the beginning of independent life (i.e.
immediately after division or birth).]

\qparr
Fontaine (F.), Stewart (E.J), Lindner (A.B.), Taddei (T.) 2008:
Mutations in two global regulators lower
individual mortality in Escherichia coli.
Molecular Microbiology  67,1,2-14.

\qparr
Gallant (J.), Palmer (L.) 1979: Error propagation in viable
cells.
Mechanisms of Ageing and Development 10,27-38.\qL
[This paper parallels at cell level the study of error 
propagation (i.e. generation after generation) undertaken
at protein level in the papers by Loftfield cited below.]

\qparr
Garvey (M.), Moriceau (B.) , Passow (U.) 2007: 
Applicability of the FDA [fluorescein diacetate, a stain]
assay to determine the 
viability of marine phytoplankton under different 
environmental conditions.
Marine Ecology Progress Series 352,17-26.\qL
[The paper contains impressive pictures of stained cells.]

\qparr
Kelly (C.D.), Rahn (O.) 1932: The gowth rate of individual
bacterial cells.
Journal of Bacteriology 23,3,147-153.

\qparr
Kibota (T.T.), Lynch (M.) 1996: Estimate of the genomic 
mutation rate deleterious to overall fitness in E. coli.
Nature 381,694-696. 
[The paper gives a death rate of 0.05 per 1,000 as
resulting from naturally occurring genetic mutations.
This is some 10 times smaller as what is observed.]

\qparr
Loftfield (R.B.) 1963: The frequency of errors in protein
biosynthesis.
Biochemical Journal 89,82-92.

\qparr
Loftfield (R.B.), Vanderjagt (D.) 1972: The frequency of 
errors in protein biosynthesis.
Biochemical Journal 128,1353-1356.\qL
[Nine years after a first article the same author published a second
article bearing the same title in which he tries an
improved methodology. Whereas the frequency of mistranslations
was estimated to be 1 part in 3,000 in the first paper,
it wa found to be 3 parts in 10,000 in the second, that is
to say basically the same result.]

\qparr
Orskow (J.) 1922: Method for the isolation of
bacteria in pure cultures from single cells and procedure for the
direct tracing of bacterial growth on a solid medium.
Journal of Bacteriology 7,537-549.

\qparr
Pierucci (O.) 1978: Dimensions of {\it Eschrichia coli} at various
growth rates: model for envelop growth.
Journal of Bacteriology 135,2,559-574.

\qparr
Berrut (S.), Pouillard (V.), Richmond (P.), Roehner (B.M.) 2016:
Deciphering infant mortality. 
Physica A 463,400-426.

\qparr
Sezonof (G.), Joseleau-Petit (D.), D'Ari (R.) 2007: 
{\it Eschrichia coli} physiology in Luria-Bertani broth.
Journal of Bacteriology December 2007,8746-8749.\qL
[The authors propose a detailed investigation of the
exponential growth phase of E. coli which, following the
lag phase, commonly lasts of the order of two hours.]

\qparr
Steiner (U.S.), Lenart (A.), Ni (M.), Chen (P.), Song (X.), 
Taddei (F.), Vaupel (J.W.), Lindner (A.B.) 2019:
Two stochastic processes shape diverse senescence patterns 
in a single-cell organism.
Evolution 73,4,847-857.

\qparr
Stewart (E.J.) Madden (R.), Paul (G.), Taddei (F.) 2005:
Aging and death in an organism that reproduces by
morphologically symmetric division.
Plos Biology 3,2,e45.

\qparr
Wilson (G.S.) 1922: The proportion of viable bacteria in
young cultures with especial reference to the technique
employed in counting.
Journal of Bacteriology 7,4,405-446.

\qparr
Woldringh (C.L.), Binnerts (J.S.), Mans (A.) 1981: Variation
in {\it Eschericia coli} buoyant density measured in Percoll
gradients.
Journal of Bacteriology 148,1,58-63.

\end{document}